\documentclass[preprint,12pt]{aastex}

\newcommand{\kms}{\mbox{km s$^{-1}$}}

\def\lsim {$\rlap{\raise.4ex\hbox{$<$}}\lower.55ex\hbox{$\sim$}\,$}

\newcommand{\lsun}{\mbox{L$_\odot$}}
\newcommand{\msun}{\mbox{M$_\odot$}}

\newcommand{\lint}{\mbox{$L_{int}$}} 

\newcommand{\andre}{Andr\'{e}}

\def\c17o{$\rm C^{17}O$}
\def\dc18o{$\rm C^{18}O$}

\begin{document}

\title {\bf Chemical Evolution in VeLLOs} 
\author {Jeong-Eun Lee}
\affil{\it Department of Astronomy and Space Science, Sejong University,
Seoul 143-747, Korea}
\email{jelee@sejong.ac.kr}

\begin{abstract}

A new type of object called ``Very Low Luminosity Objects (VeLLOs)" has been
discovered by the Spitzer Space Telescope. VeLLOs might be substellar 
objects forming by accretion. However, some VeLLOs are associated
with strong outflows, indicating the previous existence of massive 
accretion. The thermal history, which significantly affects the chemistry, 
between substellar objects with a continuous low accretion rate and objects in 
a quiescent phase after massive accretion (outburst) must be greatly different. 
In this study, the chemical evolution has been calculated in an episodic 
accretion model to show that CO and N$_2$H$^+$ have a relation different from
starless cores or Class 0/I objects. Furthermore, the CO$_2$ ice feature at
15.2 $\micron$m will be a good tracer of the thermal process in VeLLOs.   

\end{abstract}


\section{INTRODUCTION}

The \emph{Spitzer Space Telescope (SST)}, with its very high mid-infrared 
sensitivity, has revealed a new type of low-mass protostars called 
``Very Low Luminosity Objects (VeLLOs; e.g., Young et al. 2004). 
Defining the internal luminosity of a source, \lint, to be the total 
luminosity of the central protostar and circumstellar disk (if present), 
a VeLLO is defined to be an object embedded within a dense core that 
meets the criterion \lint\ $\leq$ 0.1 \lsun\ (Di Francesco et al. 2007).
VeLLOs are primarily being discovered in cores previously classified as
starless based on observations with the \emph{Infrared Astronomical 
Satellite (IRAS)} and the \emph{Infrared Space Observatory (ISO)}.
The three most well-studied VeLLOs to date are IRAM 04191 
(\andre\ et al. 1999; Dunham et al. 2006), 
L1014 (Young et al. 2004; Crapsi et al. 2005; Bourke et al. 2005; 
Huard et al. 2006), and L1521F (Bourke et al. 2006), which show very 
different properties in molecular line observations despite their
similar internal luminosities.
IRAM 04191 (\andre\ et al. 1999; Belloche et al. 2002) and L1521F 
(Crapsi et al. 2004) show evidence for infall whereas L1014 
does not (Crapsi et al. 2005).
IRAM 04191 is associated with a well-collimated outflow (\andre\ et al. 1999); 
the other two are not, although at least L1014 and possibly L1521F feature 
weak, compact outflows (Bourke et al. 2005; Bourke et al. 2006).

The discovery of VeLLOs with \emph{Spitzer} has put into question the 
picture of low-mass star formation as a continuous process of constant mass 
accretion at the standard rate of $\sim 2\times 10^{-6}$ \msun\ $yr^{-1}$ (Shu, 
Adams, \& Lizano 1987) through a single evolutionary sequence, the 
well-established class system progressing from Class 0 to III 
(Myers \& Lada 1993, Andr\'e et al. 1993).  
This standard accretion rate predicts a much higher 
luminosity than observed for VeLLOs; VeLLOs must feature some combination of 
a very low central mass and a very low accretion rate 
(e.g. Dunham et al. 2006).
If the accretion continues at the current low rate to the very small central
mass, it might not make a star. However, the accretion rate is not necessarily
constant. For instance, FU Orionis (FU Ori) objects undergo outbursts 
(Bell et al. 1995 and references therein). 
Studies (Vorobyov \& Basu 2005 and references therein) for 
the nature of the FU Ori variables suggest accretion bursts from the disk to 
the central star by the thermal instability of the disk.

Therefore, two potential explanations for the very low luminosities of VeLLOs 
are 1) proto-brown dwarfs, 
and 2) objects in a quiescent phase of the episodic accretion process.  
The former can be discriminated from the latter with studies of the chemistry 
since they involve vastly different thermal histories, which is crucial 
to the chemical evolution.
The thermal history is especially important in interactions between gas
and ice; ice evaporation and gas freeze-out from and onto grain surfaces,
respectively, depend on the dust temperature (Lee et al. 2004).  
Proto-brown dwarfs, with their very low masses, will never experience a hot 
phase, whereas the outbursts of a cycle of episodic accretion, a short time 
period when the majority of the mass is dumped onto the central protostar, 
involve significant warming of the surrounding dust.  
The quiescent states between outbursts feature much colder dust temperatures.  
As a result, envelopes of proto-brown dwarfs will be similar 
to starless cores in their chemical distributions, while objects in a 
quiescent state of episodic accretion will show different chemical 
distributions from starless cores or normal, embedded Class 0/I objects. 

IRAM 04191 may be undergoing episodic accretion since it features a 
strong outflow which predicts a higher accretion rate by two orders of 
magnitude than inferred from the internal luminosity of the source 
(\andre\ et al. 1999; Dunham et al. 2006).
Furthermore, the N$_2$H$^+$ emission, observed with 
the Plateau de Bure Interferometer (PDBI) and the IRAM 30 m telescope, 
shows a hole in the center of the envelope (Belloche \& \andre\ 2004).
In general, N$_2$H$^+$ emission tends to peak towards the center of 
starless cores (Lee et al. 2003), but be deficient from the centers of
Class 0/I sources due to destruction by CO as it evaporates 
(Lee et al. 2004).
IRAM 04191 shows moderate CO depletion (Crapsi et al. 2004); 
Belloche \& \andre\ (2004) suggest that freeze-out of N$_2$ in the 
high-density, inner envelope might result in the observed N$_2$H$^+$ 
hole.  
However, if freeze-out of N$_2$ is significant enough to explain this hole, 
there should be significantly more deuteration and depletion of CO than
observed (Crapsi et al. 2004), similar to that seen in 
prestellar cores (Lee et al. 2003). 

In this study, we model the chemical evolution in the process of episodic
accretion to provide a possible explanation of the chemical distributions
of CO and N$_2$H$^+$ in gas and to predict observable consequences in the 
CO$_2$ ice feature in VeLLOs such as IRAM 04191 that show 
strong evidence for undergoing such a process.

\section{EPISODIC ACCRETION MODELS}

We use the chemo-dynamical model developed by Lee et al. (2004). 
This model calculates the chemical evolution of a model core
evolving from the prestellar stage through the embedded 
protostellar stages.  
The dynamical evolution is described by combining a sequence of Bonnor-Ebert 
spheres (Bonnor 1956, Ebert 1955) with the inside-out collapse 
model (Shu 1977), where the accretion rate from the envelope onto the star+disk 
system is constant. 
The model also includes the First Hydrostatic Core (FHSC) stage, which
results from the first gravitational collapse of a dense molecular core and 
lasts until the core temperature reaches 2000 K and the 
dissociation of molecular hydrogen causes the second collapse 
(Boss 1993, Masunaga et al. 1998).
The radius of the FHSC is about 5 AU, and the disk is not yet well developed 
at this stage.
As a result, the accretion luminosity arising from accretion onto the 
FHSC is not significant.
The evolution of the central luminosity follows that adopted by 
Young \& Evans (2005), who incorporated the evolution of an unresolved 
disk and stellar photosphere into the central luminosity.
This luminosity is proportional to the accretion luminosity arising 
from envelope accretion following the 
inside-out collapse model (see Young \& Evans 2005 for details).

In this study, we assume that the accretion from the envelope to the disk is 
constant, but the accretion from the disk to the star is episodic.
Thus, the envelope density structure is identical to that of the inside-out 
collapse model, but the internal luminosity is changed by the episodic 
accretion process.  
The initial mass of the model core, with an effective sound speed of 0.27 \kms\ 
and outer radius of 6200 AU, is 1 \msun\ (Equation 1 of Young \& Evans 2005). 
The episodic accretion is assumed to begin after the FHSC phase, which lasts 
for $2\times 10^4$ years.

The dust temperature profile at each time step is calculated with the one 
dimensional continuum radiative transfer code DUSTY (Ivezi\'c et al. 1999), 
using as input the density profile from the inside-out collapse model. 
The input internal radiation field at each time step is calculated by the model 
of Young \& Evans (2005) with their standard parameters of $\eta_D = 0.75$, 
$\eta_* = 0.5$, and $\tau_{max}=10$ (see Young \& Evans 2005 for a detailed 
explanation of these parameters) combined with our prescription for
episodic accretion.  In this prescription, accretion bursts 10 times more 
massive than normal ($\dot{M}\sim 5\times 10^{-6}$ M$_\odot$ yr$^{-1}$) occur 
during $10^3$ years every $10^4$ years.
For the $9\times 10^3$ years between the accretion burst events,
the accretion from the disk to the star is decreased by a factor of 100.
Vorobyov \& Basu (2005) have showed that $0.01\sim 0.05$ M$_\odot$ is accreted
from the disk to the central source with the accretion rate of $(1\sim 10) 
\times 10^{-4}$ M$_\odot$ yr$^{-1}$ during a burst, 
and the duration of the intervening quiescent accretion phase is $(1\sim 3) 
\times 10^{3}$ years for $\dot{M}=(1-10)\times 10^{-7}$ M$_\odot$ yr$^{-1}$.
According to their calculations and for the accretion rates in the burst and 
quiescent phases in this model, which are about an order of 
magnitude smaller than the rates of Vorobyov \& Basu (2005), we adopt the 
timescales of $10^3$ and $10^4$ years in the two phases, respectively, and 
these timescales are consistent with the estimated from HL Tauri by 
Lin et al. (1994).

We calculate the radiative transfer independently at each time step since 
the dust temperature reaches a new equilibrium essentially instantly compared 
to the length of our time step.  For example, one silicate grain with 
a radius of 0.1 \micron\ and a temperature of 100 K will cool down to 10 K 
within one second if the heating source suddenly disappears 
(Draine \& Anderson 1985).  
Following Young \& Evans (2005), the envelope is 
heated externally by the Black-Draine Interstellar Radiation Field (ISRF) 
(Evans et al. 2001) attenuated by a visual extinction of 0.5 mag.  
Also following Young \& Evans, we used the dust model of Ossenkopf \& Henning
(1994) appropriate for thin ice mantles after $10^5$ years of coagulation 
at a gas density of $10^6$ cm$^{-3}$ (OH5 dust).

Fig. 1a shows the evolution of the total internal luminosity in the model, 
and Fig. 1b shows the temperature evolution of a few infalling grid points.  
As seen in this figure, the temperature that an infalling gas parcel experiences 
peaks at the time step of each massive accretion and is lowered significantly 
during the quiescent states.
Betwen 4 and 5.5$\times 10^4$ years, the luminosity and temperature increase 
steeply with time even in the quiescent phase because of the transition from 
the FHSC to Class 0, which occurs from 2$\times 10^4$ years until 
5.5$\times 10^4$ years. 
This transition timescale is the same as that used in Lee et al. (2004),
but much longer than the 100 years assumed by Young \& Evans (2005).

The evolution of the spectral energy distribution (SED) around $5\times 10^4$
years is presented in Fig. 2.  
An accretion burst occurred at $5.1\times 10^4$ years, after which
a quiescent state continues until the next burst $10^4$ years later.  
The accretion luminosity powered in the burst phase at $5.1\times 10^4$ years is
about 5 \lsun, whereas the accretion luminosities at 5.0 and 5.4 $\times 10^4$ 
years are about 0.004 and 0.02 \lsun, respectively.  
The model SED at $5.4\times 10^4$ years fits well the observed SED of 
IRAM 04191.  
The total luminosity in the model is about 0.09 \lsun, where the 
luminosity arising from heating by the ISRF ($\sim 0.07$ \lsun) is 
larger than the internal luminosity ($\sim 0.02$ \lsun).  
This result is not very consistent with that of Dunham et al. (2006), 
who found that the internal luminosity must be greater than 0.04 \lsun.  
However, the density profile and ISRF in their model are 
significantly different from ours, making it hard to directly compare the 
two results.

The density and dust temperature profiles are shown in Fig. 3 
at the same time steps shown for the SEDs in Fig. 2.  
The infall radius is 0.014 pc (2800 AU) when the accretion burst occurs 
at $5.1\times 10^4$.
The density at the infall radius is about $2\times 10^5$ cm$^{-3}$.
The density profile has the power law index of $\sim 1.5$ and 2.0 inside
and outside the infall radius, respectively. 
Except for the accretion burst phase, the dust temperature is below the CO
evaporation temperature ($\sim 20$ K) at almost all radii.

\section{CHEMICAL EVOLUTION IN A CYCLE OF EPISODIC ACCRETION}

In prestellar cores, which are very evolved starless cores, carbon-bearing
molecules such as CS and CO are readily frozen-out onto grain surfaces 
(Bergin \& Langer 1997, Aikawa et al. 2001, Lee et al. 2003, 
Lee et al. 2004). 
In contrast, nitrogen-bearing molecules such as N$_2$H$^+$ and NH$_3$ are 
considered good tracers of (column) density in cold, dense cores since
N$_2$, the parent molecule of N$_2$H$^+$ and NH$_3$, has a long formation
timescale such that it becomes abundant later in the core evolution.
Therefore, the comparison between CO and N$_2$H$^+$ abundances is a good
indicator of the physical properties of a dense core. 

The low binding energy of N$_2$ on grain surfaces has also been suggested 
as an explanation for the long-lived molecules in the cold regions 
(Lee et al. 2004). 
However, \"Oberg et al. (2005) showed that the binding energy of N$_2$ on grain
surfaces is not different from that of CO.
The high abundance of N$_2$H$^+$ in the cold, dense regions may be due to the
replenishment of N$_2$ through the interactions between N$_2$H$^+$ and grains 
since N$_2$H$^+$ recombines with electrons on the grain surfaces.  
The significant depletion of CO and electrons, the principal 
destroyers of N$_2$H$^+$ in the gas phase, is another explanation for 
the high abundance of N$_2$H$^+$ in these cold, dense regions.

We updated the chemical network used in Lee et al. (2004) to include 
recent results on the N$_2$H$^+$ chemistry, including a new binding energy 
of N$_2$ identical to that of CO (\"Oberg et al. 2005) and new rates of
dissociative recombination of N$_2$H$^+$ with electrons (Geppert et al.  2004).  
The initial abundances of other species are the same as those in Table 3 of 
Lee et al. (2004).
Bare silicate grains were assumed for binding energies of molecules. 

Fig. 4 presents the evolution of the CO abundance in gas and ice at the time 
steps $10^3$ years before, during, and 3 and $5\times 10^3$ years after the 
accretion burst at $5.1\times 10^4$ years since the model SED at 
$5.4\times 10^4$ years fits the observations of IRAM 04191 reasonably well.
$10^3$ years before the accretion burst, CO is frozen-out onto grain 
surfaces and depleted from the gas at most radii except around the 
CO evaporation radius, inside which the dust temperature is higher than the 
CO evaporation temperature, developed by the previous 
accretion burst at $4.1\times 10^4$ years.
During the burst phase, the CO evaporation radius jumps to a much 
larger radius due to the significant increase in accretion luminosity. 
Once the burst phase ends and the core makes the transition back to the 
quiescent phase, the accretion luminosity, and therefore the temperature, 
decreases greatly.  
CO is then gradually frozen back out of the gas and onto the grain surfaces.

The evolution of the N$_2$H$^+$ abundance is shown in Fig. 5.
Before the accretion burst at $5.1\times 10^4$ years, 
the N$_2$H$^+$ abundance peaks at a rather large radius and decreases inward. 
(The sharp drop of the N$_2$H$^+$ abundance at the edge of the model 
core results from the dissociative recombination of N$_2$H$^+$ with electrons.) 
At radii of 0.002 pc (log r$\sim -2.7$) and 0.0008 pc (log r$\sim -3.1$),
the bumps of N$_2$H$^+$ are caused by the dips of CO at these radii.
The N$_2$H$^+$ abundance decreases at radii smaller than 0.0008 pc even though 
CO is depleted here.  
N$_2$H$^+$ forms mainly by the reaction between N$_2$ and H$_3^+$, which is
destroyed by CO. As mentioned above, the evaporation temperature of N$_2$ is
the same as that of CO, i.e., $\sim$20 K. 
Therefore, at the radii smaller than 0.0008 pc where the dust temperature is 
lower than $\sim 20$ K, and the density is greater than $10^6$ cm$^{-3}$, 
the significant freeze-out of N$_2$ causes the low abundance of N$_2$H$^+$.
During the burst phase, the N$_2$H$^+$ abundance drops
significantly at radii smaller than the CO evaporation radius since CO 
destroys N$_2$H$^+$. 
In the quiescent states 3, 5, and $8\times 10^3$ years after the burst, 
N$_2$H$^+$ slowly increases at small radii as CO is frozen-out onto the 
grain surfaces.

\section{DISCUSSION}

Most VeLLOs were classified as starless core before \emph{Spitzer} observations, 
and they showed wide variation in chemical and physical properties.  
For instance, IRAM 04191 (\andre\ et al. 1999) and L673-7 (M. Dunham 2007, 
private communication) have well-collimated molecular outflows, 
but no molecular outflow has been detected in L1521F even though its 
\emph{Spitzer} IRAC 
images show a compact bipolar scattered emission nebula (Bourke et al. 2006). 
L1014 features a very compact outflow detected with the Submillimeter Array
(Bourke 2005).  
In chemical properties, IRAM 04191 and L1014 are similar with moderate CO 
depletion but different in their N$_2$H$^+$ emission, which peaks at the 
center of L1014 but shows a hole at the center of IRAM04191.
L1521F, which was believed to be a highly evolved starless core (Crapsi et al. 
2004), shows significant CO depletion and centrally peaked N$_2$H$^+$ emission.   
Before the \emph{Spitzer} observations, this variation in chemical properties was 
believed to arise from an as-of-yet incompletely understood evolutionary 
sequence of starless cores.
The discovery of the VeLLOs has cast doubt on our understanding of the 
continuous and unique sequence of star formation consisting of the
evolution of a starless core into a prestellar core followed by the formation of 
a Class 0 object that then evolves through the Class system.

The sources without strong outflows might be in the final accretion phase of 
substellar objects or in the FHSC stage (Omukai et al. 2007).  
On the other hand, VeLLOs with strong, well-collimated outflows such as IRAM 04191
are hard to understand since the outflow suggests a much higher accretion luminosity than the observed internal luminosity of the source.
A possible explanation for such objects is episodic accretion.  
In this picture, VeLLOs are objects in quiescent phases of a cycle of episodic 
mass accretion.  The strong outflows observed can be generated during accretion 
bursts that occurred prior to the current quiescent phase.
These two processes predict vastly different thermal histories, which can 
be discriminated by their different predictions for the chemical 
evolution of the core.
Observations probing these different predictions are challenging, however, 
since high spatial resolution is required 
to resolve the regions most affected by the different thermal histories.  
In addition, 
observations of ice features require high sensitivities due to the very
low luminosities of the central source.

As seen in the previous section, the relation between CO and N$_2$H$^+$
abundances in an object featuring episodic accretion is different from that 
of starless cores or hot corinos, 
where CO and N$_2$H$^+$ have an anti-correlation.  
In the quiescent phase after an outburst, both CO and N$_2$H$^+$ have not 
had enough time to reach chemical equilibrium.
As a result, CO can be less depleted and N$_2$H$^+$ more depleted 
compared to a core forming a proto-brown dwarf with a similar central 
luminosity to IRAM 04191.

CO in ice can also be a discriminator since CO ice features
from VeLLOs in a quiescent phase of episodic accretion will be weaker  
than those in proto-brown dwarfs.
Additionally, the CO$_2$ ice features can also be very 
important tracers of the thermal history of VeLLOs.
The CO$_2$ 15.2 $\micron$ bending mode has been detected toward massive star 
forming regions (Gerakines et al. 1999),
low mass star forming cores (Pontoppidan et al. 2007)
and even dark molecular clouds (Knez et al. 2005).
The formation process of the CO$_2$ ice is controversial, but it must form on 
grain surfaces since the abundance of the CO$_2$ gas in dark clouds is too low 
($10^{-9}$, Bergin et al. 1995) to explain the observed CO$_2$ ice features.
CO$_2$ ices are found both in water- and CO-rich grain surfaces.
Therefore, the CO$_2$ ice likely forms on grain surfaces by the reactions 
between CO frozen-out from gas and OH or O photodissociated from water or CO
in ice.
The photodissociation of water or CO on grain surfaces, especially in
low mas star forming cores, can occur by 
the cosmic ray induced UV photons. As a result, the timescale of the formation
of CO$_2$ on grain surfaces must be long ($ > 10^4$ years in the density of 
$10^5$ cm$^{-3}$, Pontoppidan et al. 2007.) 

The evaporation temperature of the CO$_2$ ice is about 35 K, higher and
lower than the CO (20 K) and H$_2$O (100 K) evaporation temperatures, 
respectively.
In outburst phases of episodic accretion, 
the dust temperature can rise above the CO$_2$ evaporation temperature at some
inner radii. There is also a range of radii where the dust temperature is 
between the CO and CO$_2$ evaporation temperatures, so CO$_2$ remains on 
grain surfaces while CO evaporates. In this range, CO$_2$
will produce the double-peaked feature indicative of pure CO$_2$ ice 
(Pontoppidan et al. 2007). 
In the following quiescent phase, CO will be gradually frozen-out as seen
above, and thus, combined with the long formation timescale of CO$_2$, the 
solid CO$_2$ abundance profile developed in the outburst phase will not change 
greatly. 
As a result, we expect that objects in quiescent stages of episodic 
accretion will have a weaker solid CO$_2$ feature (if CO$_2$ evaporates up to
a large radius) but a stronger pure CO$_2$ feature (if the range of 
CO$_2$ evaporation is not large) compared to substellar objects. 
Fig. 6 shows the temperature profile during an outburst 
with marks for the CO and CO$_2$ ice mixture.
However, if the VeLLOs are proto-brown dwarfs, which never raise the dust
temperature above the CO and CO$_2$ evaporation temperatures in the majority of 
their envelopes, 
then the solid CO$_2$ feature will be strong as seen toward
background sources without a strong pure CO$_2$ component (Knez et al. 2005). 
Therefore, the CO$_2$ ice feature as well as the abundance distribution of
CO and N$_2$H$^+$ in gas can be good tracers of the formation processes of 
VeLLOs.

\acknowledgments

This work was supported by the faculty research fund of Sejong
University in 2007.
We are very grateful to Mike Dunham and Neal Evans for valuable comments.

\clearpage

\begin{figure}
\figurenum{1.a}
\epsscale{1.0}
\plotone{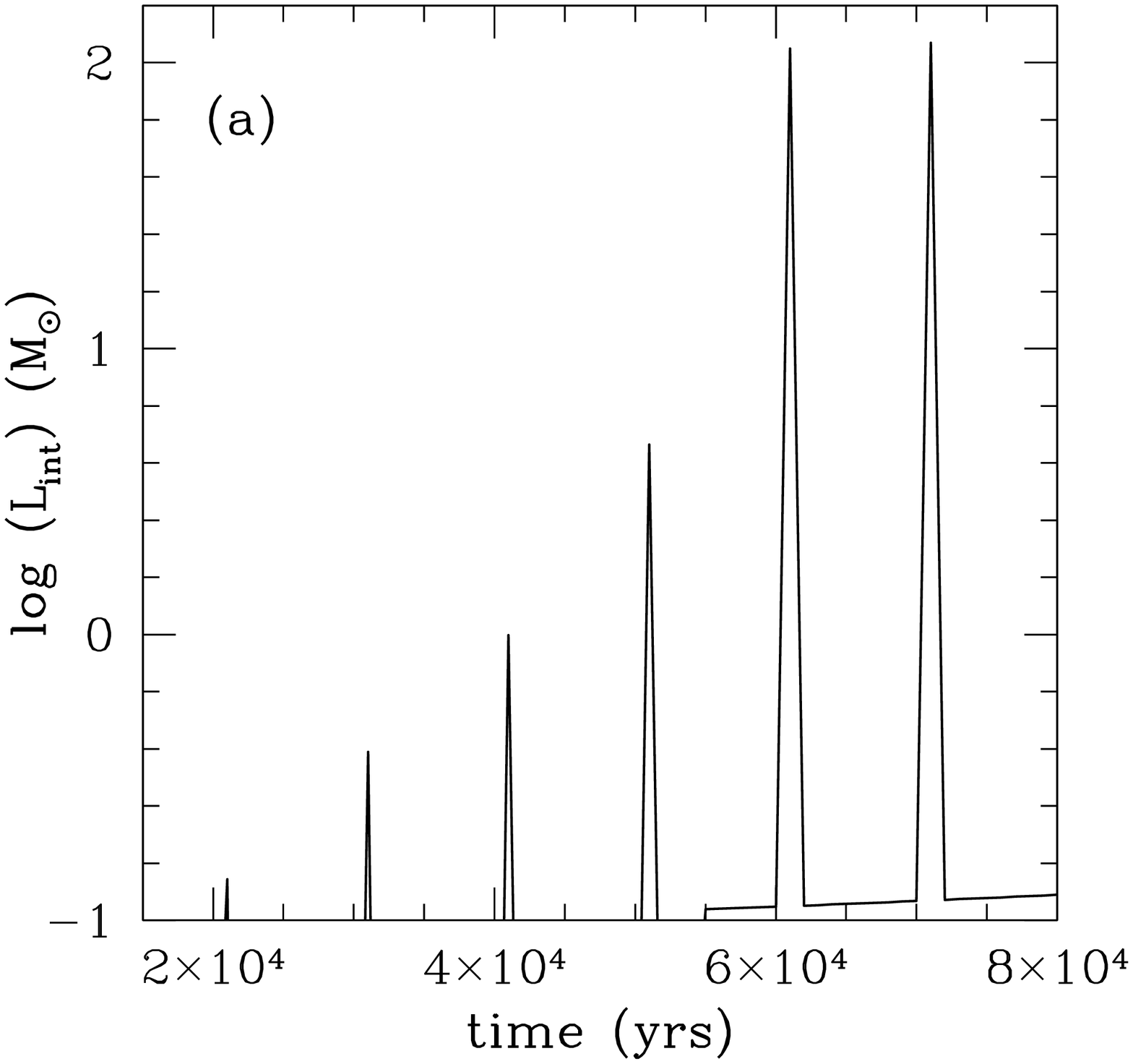}
\caption{Evolution of the internal luminosity in the episodic accretion model.
Outburts (massive accretion events) occur every $10^4$ years for $10^3$ years
after the FHSC phase, which lasts until $2\times 10^4$ years.
}
\end{figure}

\clearpage

\begin{figure}
\figurenum{1.b}
\epsscale{1.0}
\plotone{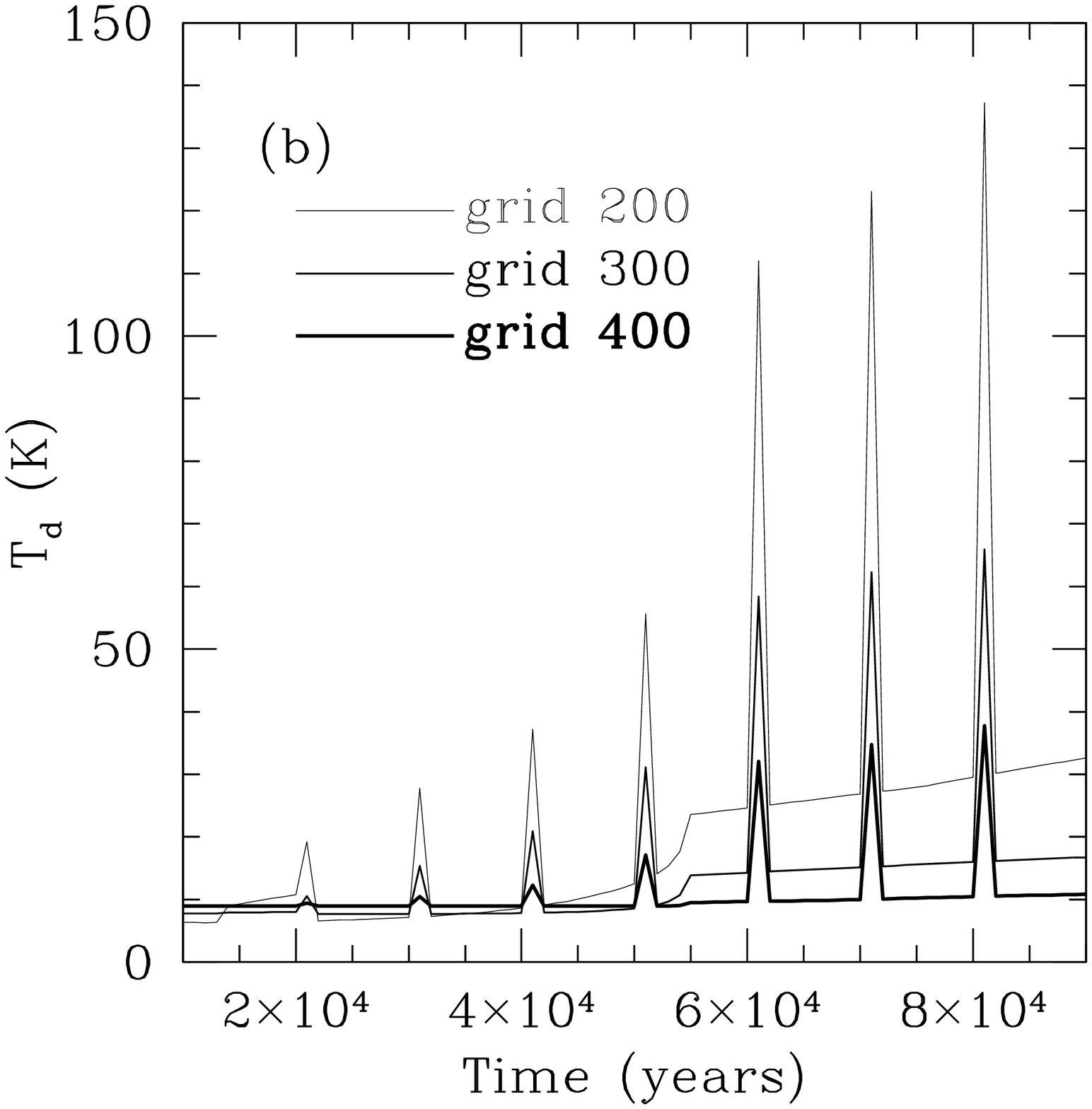}
\caption{Evolution of dust temperatures in infalling gas parcels.
Grids 200, 300, and 400 are initially located at the radii of 
0.0023, 0.0055, and 0.0124 pc, respectively.
}
\end{figure}

\clearpage

\begin{figure}
\figurenum{2}
\epsscale{1.0}
\plotone{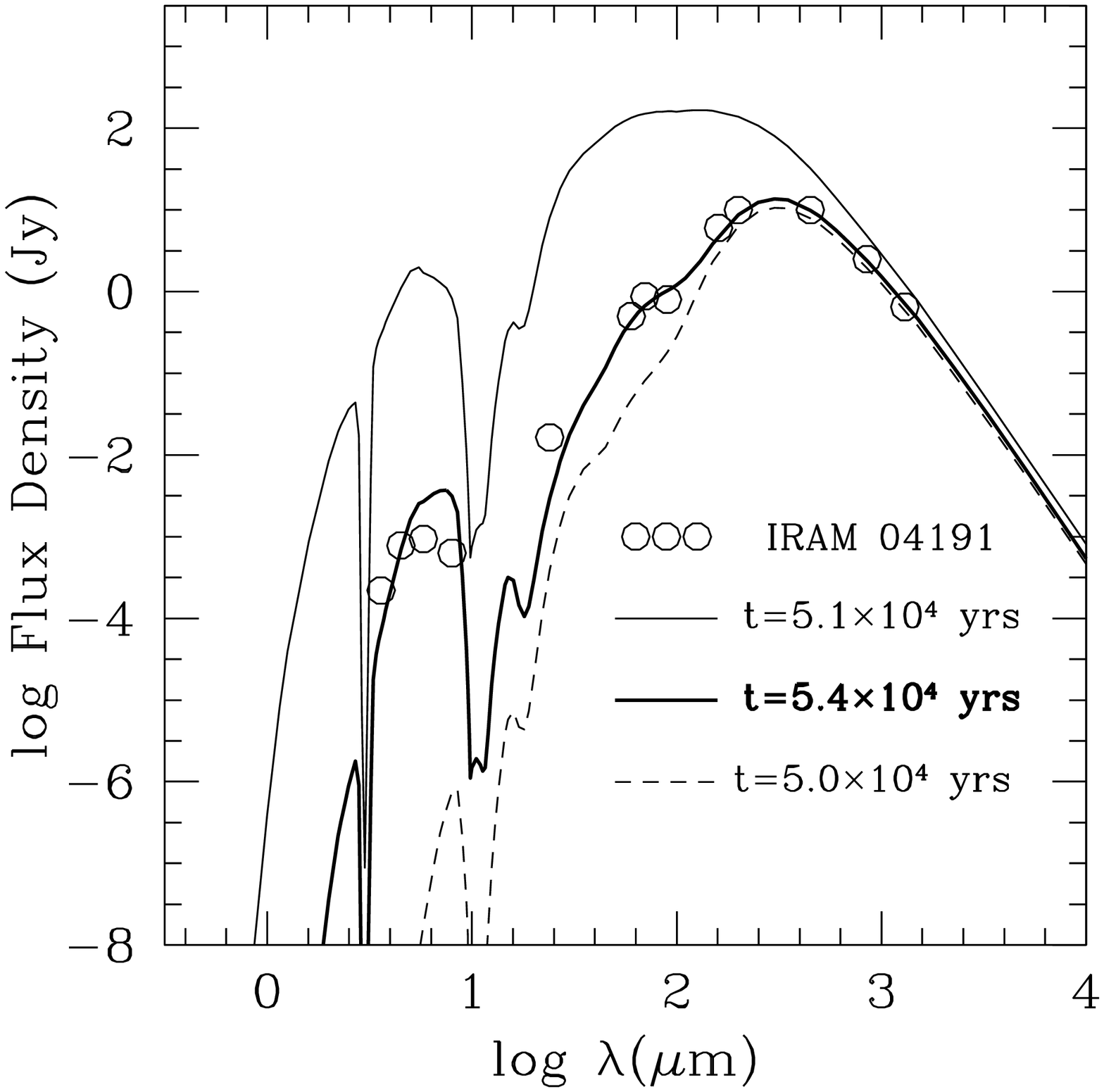}
\caption{Evolution of SED in the episodic accretion model.
The dashed, thick solid, and thin solid lines represent the SEDs 
$10^3$ years before, $3\times 10^3$ years after, and during the 
outburst at $5.1\times 10^3$ years, respectively.
}
\end{figure}

\clearpage

\begin{figure}
\figurenum{3}
\epsscale{1.0}
\plotone{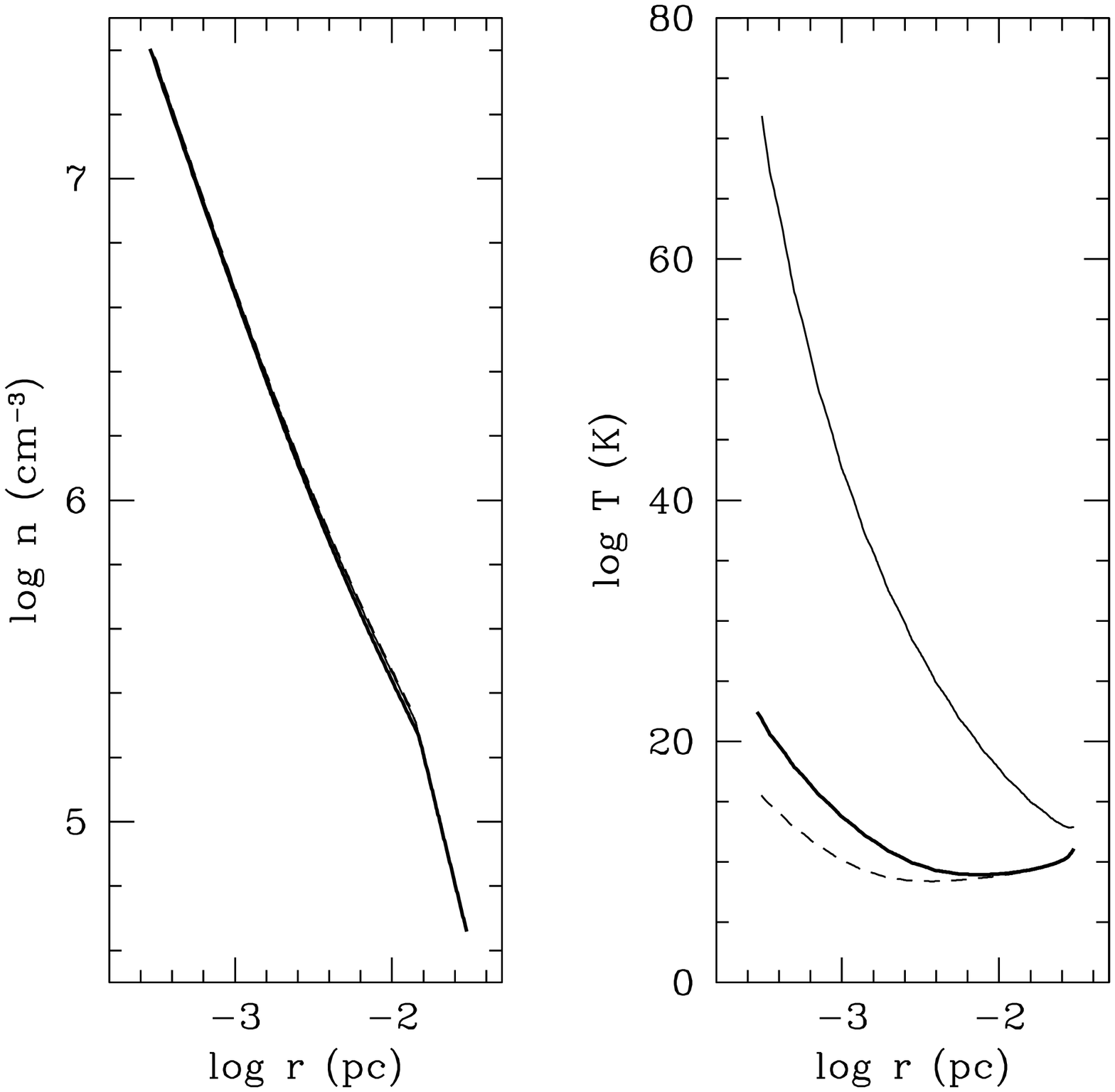}
\caption{Evolution of density and temperature profiles.
Time steps for different line styles are the same as those in Fig. 2.
The density structure in the envelope does not change much for 
$4\times 10^3$ years, but the dust temperature varies significantly 
over the episodic accretion event.
During the outburst, the dust temperature jumps above the CO evaporation
temperature at radii smaller than about 0.006 pc ($\sim$ 1300 AU).
During the quiescent states, the dust temperature is below the CO evaporation
temperature ($\sim 20$ K).
}
\end{figure}

\clearpage

\begin{figure}
\figurenum{4}
\epsscale{1.0}
\plotone{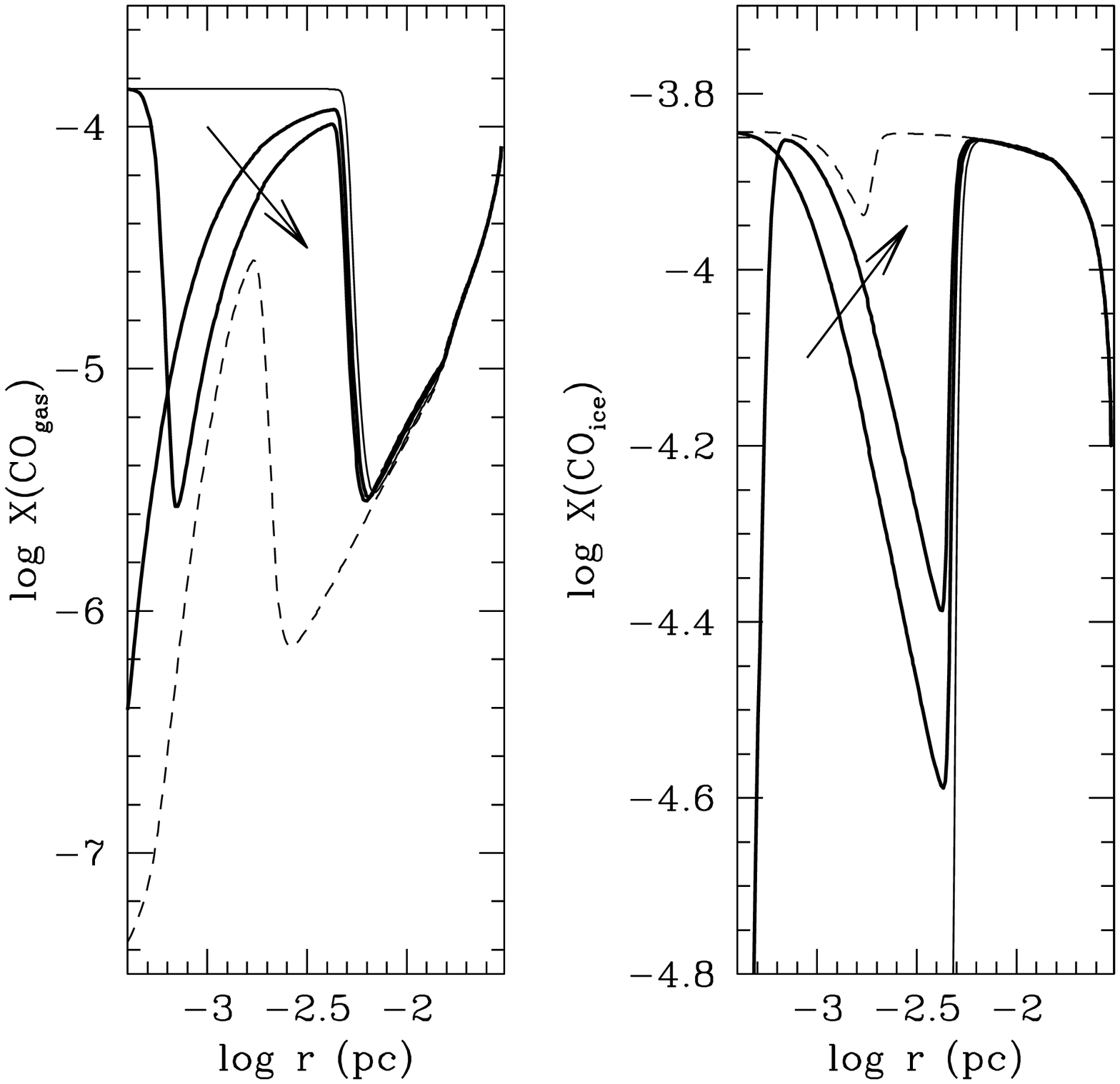}
\caption{ Evolution of the abundance profile of CO in gas and ice.
Time steps for different line styles are the same as those in Fig. 2 except
for one more thick solid line $5\times 10^3$ years after the outburst event.
Before the outburst (dashed line), the CO abundance profiles in gas and ice
have a peak and a dip, respectively, at $\sim$ 0.002 pc, which are vestiges of the
prior outburst at $4.1\times 10^4$ years.
During the outburst (thin solid line), CO evaporates, and thus,
the CO abundance in gas reaches the maximum at radii smaller than 0.006 pc.
For the CO ice, the opposite happens, i.e., the abundance drops sharply at 
0.006 pc.
The arrow in each panel indicates the evolution of the 
abundance profile with time after the massive accretion event.
That is, CO is gradually frozen-out onto grain surfaces to become depleted in gas 
and abundant in ice with time.
}
\end{figure}

\clearpage

\begin{figure}
\figurenum{5}
\epsscale{1.0}
\plotone{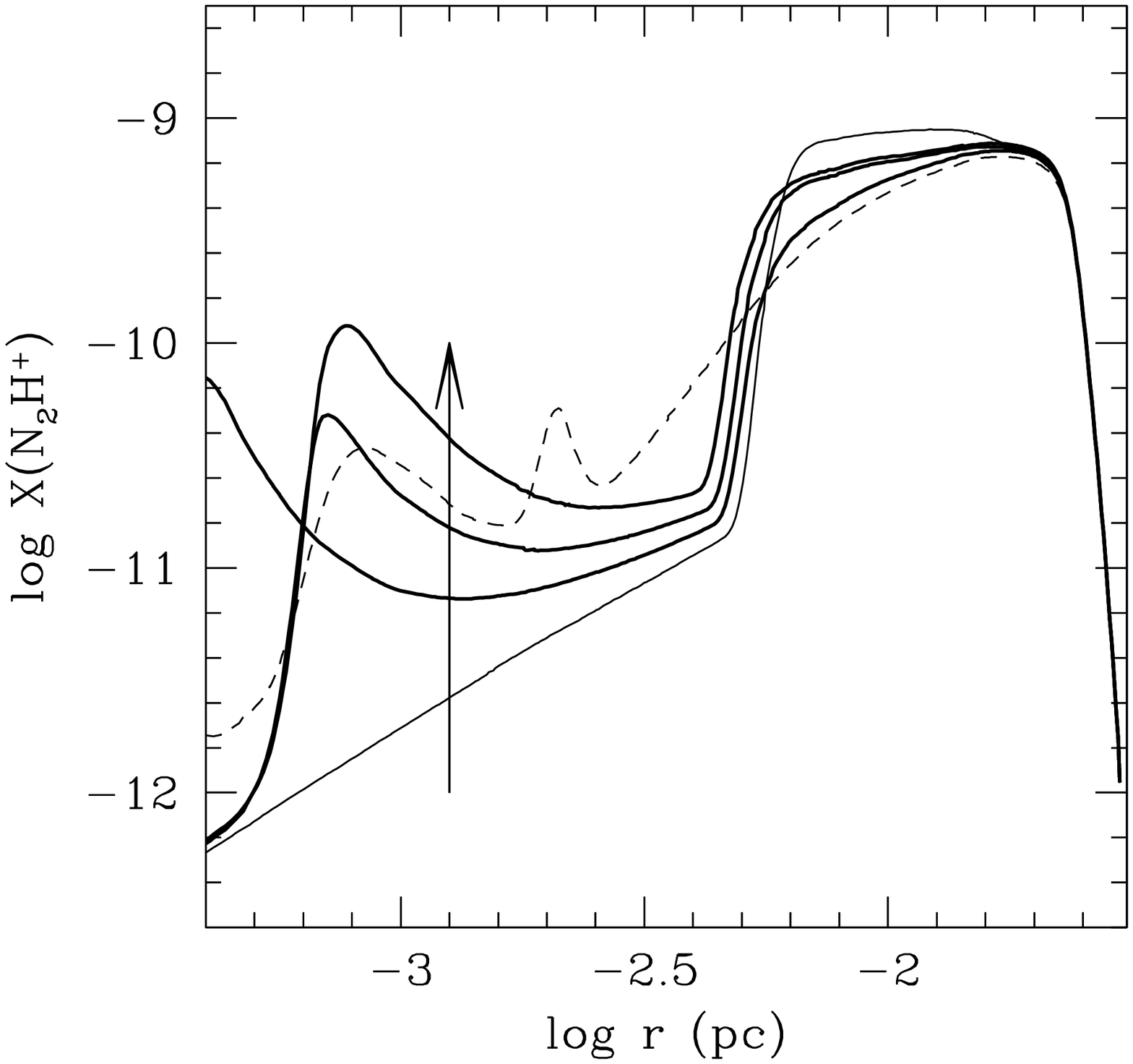}
\caption{Evolution of the abundance profile of N$_2$H$^+$.
The time steps for different line types are the same as those in previous figures,
but the thick solid lines show the abundance evolution 3, 5, and 8 $\times 10^3$ 
years after the outburst as indicated with the arrow.  
The bumps shown in the dashed line (before the outburst) are vestiges of the 
previous outburst that occurred at $4.1\times 10^4$ years.
During the outburst (thin solid line), N$_2$H$^+$ is destroyed by CO inside the CO evaporation radius, so its abundance drops sharply at radii smaller than the CO 
evaporation radius. 
In the quiescent phase (thick solid lines), the N$_2$H$^+$ abundance gradually 
increases as CO is frozen out onto grain surfaces.
}
\end{figure}

\clearpage

\begin{figure}
\figurenum{6}
\epsscale{1.0}
\plotone{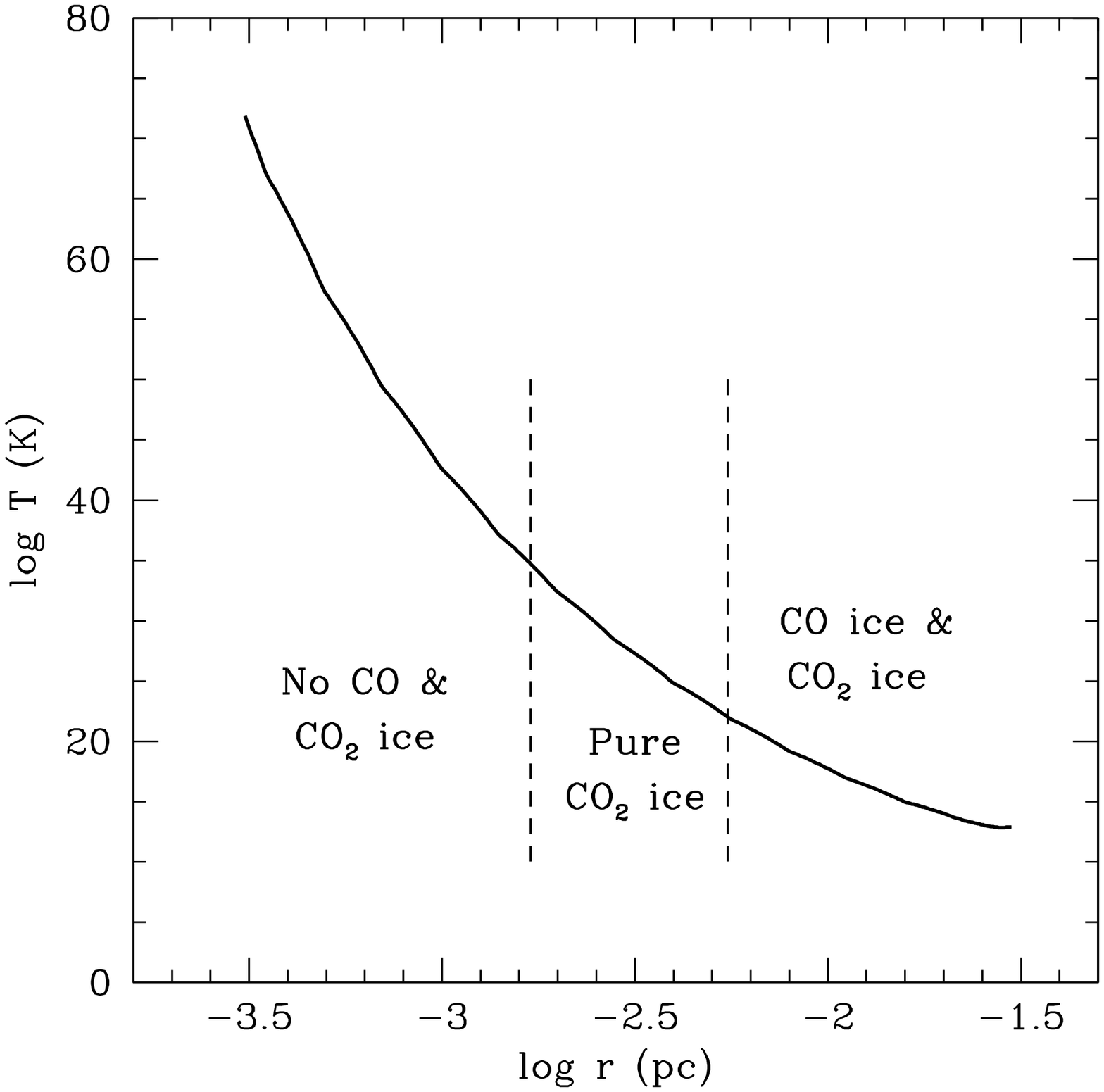}
\caption{Distribution of CO and CO$_2$ ices depending on the temperature 
distribution developed by the outburst at $5.1\times 10^4$ years.
}
\end{figure}

\end{document}